\documentclass[preprint,amsmath,amssymb,aps,prd,showpacs]{revtex4-1}

\usepackage{epsfig}
\usepackage{dcolumn}
\usepackage{bm}
\usepackage{tikz}
\usepackage{graphicx, graphics, color}
\usepackage{dcolumn}
\usepackage{bm}
\usepackage{hyperref}
\usepackage[mathlines]{lineno}
\usepackage{float}
\usepackage{subfig}

\newcommand{\be}{\begin{equation}}
\newcommand{\ee}{\end{equation}}
\newcommand{\ben}{\begin{eqnarray}}
\newcommand{\een}{\end{eqnarray}}
\newcommand{\la}{{\lambda}}

\newcommand{\p}{\partial}
\newcommand{\na}{\nabla}

\newcommand{\tA}{\tilde A}
\newcommand{\tF}{\tilde F}

\newcommand{\ep}{\epsilon}

\newcommand{\tB}{{\tilde B}}

\newcommand{\te}{\tilde e}
\newcommand{\tj}{\tilde j}

\graphicspath{{figs/}}
\graphicspath{{../figs/}}

\begin{document} 

\title{Chiral kinetic theory in the presence of dark photon}

\author{Marek Rogatko} 
\email{rogat@kft.umcs.lublin.pl}
\author{Karol I. Wysokinski}
\email{karol@tytan.umcs.lublin.pl}
\affiliation{Institute of Physics, 
Maria Curie-Sklodowska University, 
pl.~Marii Curie-Sklodowskiej 1,  20-031 Lublin,  Poland}


\date{\today}

\begin{abstract}
Finding signatures of dark matter 
in transport characteristics of solids would be an important step on the road to detect this illusive component of the mass of our Universe. This is especially important 
and timely as the experiments designed to directly detect dark matter particles continue to provide negative results. As a first step in this direction we consider 
topologically non-trivial Weyl or Dirac semimetals and derive 
the modified kinetic equation taking into account two coupled $U(1)$-gauge fields, one being the standard Maxwell electromagnetic field and other corresponding to the dark sector. The resulting Boltzmann kinetic equation is modified by the Berry curvature which couples to both visible and dark sector gauge fields. It was revealed that the dark sector induces  modifications of transport coefficients due to the appearance of coupling constant between gauge fields and dark sector magnetic field. 

\end{abstract}

\maketitle
\flushbottom

\section{Introduction}
\label{sec:intro}

The wealth of gravitational evidence for the existence of {\it dark matter} contrasts with the absence of direct terrestrial observations of this abundant component of mass of the Universe~\cite{bergstorm2000,bertone2005,freese2013}. 
The {\it dark matter} interactions with baryonic matter is extremely weak or even null. However its existence has been deduced from analysis of gravitational effects, such as
galactic rotation curves, gravitational lensing, the large-scale structure formation of the Universe and the cosmic microwave background radiation.

Many experiments aimed at the direct detection of dark matter include approaches relying on the occasional interactions of visible and {\it dark matter} particles. New types of fundamental particles, being claimed as candidates for {\it dark matter } sector, potentially
ought to interact with nuclei in detector materials on Earth. However, only DAMA collaboration \cite{ber98,ber13} announced the observed modulation in the rate of interaction events, which may constitute the spur of {\it dark matter} evidence. But extraordinary claims require extraordinary evidence. Several groups want to reproduce the DAMA results but in vain \cite{cos18}.

On the other hand, the very recent reports from two experiments indicate negative search results \cite{bartram2021,meng2021} (in the elaborated mass range) for WIMPs and axions, being
the main candidates for {\it dark matter} particles due to their potential to simultaneously solve other problems in the Standard Model. The only possible signal of direct {\it dark matter} detection remains that from XENON1T \cite{aprile2020}, the team still  works to achieve  better statistics.

However because of the growing sense of some kind of a {\it crisis} in the {\it dark matter} search for, stemming from the absence of the evidence for the most popular candidates for {\it dark sector} particles, some new diversifying experiments are paid attention to. One ought to look for the 
unconventional experiments and techniques of detection \cite{ber18}.

One of the directions 
is related to the implementation of molecular or condensed matter systems 
including superconductors or superconducting devices. 
Recent proposals comprise the search for bosonic dark matter {\it via} absorption 
in superconductors \cite{hochberg2016}, using superfluid helium \cite{knapen2017} or  even
optical phonons in polar materials \cite{knapen2018}. The others,
are based on the observations of color centers production in crystals \cite{budnik2018}, 
or the usage of bulk three-dimensional Dirac semimetals \cite{hochberg2018}, topological 
semiconducting compounds \cite{liang2018}, and
multilayered optical devices \cite{bar18}.

Direct methods to detect {\it dark matter} particles propose to use different target  materials with suitable characteristics as discussed in numerous reviews
\cite{bertone2005,mayet2016,gaskins2016}. 
The choice of the target depends on the mass range of expected {\it dark sector} particles and the energy deposited during the scattering with ordinary matter. Recently, the Dirac semimetals 
have been proposed~\cite{hochberg2018} as efficient targets for detection of very light {\it dark matter}, with masses in the keV to MeV range. 
These materials possess linear energy spectrum with no or very tiny gap (when some symmetries are broken) between valence and conduction band and may thus be sensitive to very low energies deposited during scattering of {\it dark matter} with electrons. Similar spectrum possess Weyl semimetals with even number of Dirac cones of various chiralities of massless quasiparticles residing in symmetry related cones. These systems are known to break time reversal symmetry and exhibit anomalous Hall effect~\cite{haldane2004} in absence of  any magnetic field. In the quasi-classical description~\cite{xiao2010} this behaviour is related to the dynamics of charged particles subject to the Berry curvature which  in the nontrivial way modifies the Boltzmann kinetic equation.  

The most recent review of experimental developments and ideas of the quest for {\it light dark matter} from keV to sub-GeV, with the implementation of condensed matter systems, has been
published in \cite{kah22}.

 In this context one should mention recent proposals of detecting axionic {\it dark matter}~\cite{iwazaki2020,iwazaki2021} in condensed matter systems, namely metals and superconductors. They rely on the appearance of the high frequency surface electric field if  the axions are subject to the external magnetic field. The idea bases on the chirality of the axions which are coupled to both electric and magnetic fields. The resulting high frequency electric field emerging on the surface of metals or superconductors~\cite{kishimoto2022} can possibly be detected.

Another approach of {\it dark matter} detection utilises the model of {\it dark photon}, pertaining to the class of Weakly Interacting Sub-eV Particles (WISPs), with
expected kinetic mixing between Maxwell electromagnetic field and the auxiliary $U(1)$-gauge field connected with the 
{\it dark sector}. From the gauge theory point of view, the model can be regarded as
an extension of the Standard Model, by the introduction of the auxiliary hidden sector $U(1)$ symmetry. On the other hand, 
at the level of the unification theories, one can think about the model as, emerging by compactification the different models of string/M-theory which induces the new $U(1)$-gauge groups.
It can be accounted as {\it dark matter} sector.

In this attitude one assumes the electromagnetic background to consist of two $U(1)$-gauge field strengths, denoted as 
$F_{\mu \nu}$ (Maxwell) and $B_{\mu \nu}$ ({\it dark sector}), 
coupled together as $\alpha F_{\mu \nu} B_{\mu \nu}$, constituting the so-called kinetic mixing term.

This mixing modifies various properties of the condensed matter systems as, e.g., the superconducting transition temperature $T_c$ which gets modified from its bare value $T_c(0)$ to the actual value $T_c(\alpha)$.  The problems in question were widely treated in the holographic attitude to the {\it dark matter} sector and its influence on various aspects of physical properties
of superconductors/superfluids and vortices \cite{nak14}-\cite{pen15b}. In the aforementioned 
 the so-called holographic
approach, {\it dark sector} emerges from the top-down reduction of superstring/M-theory.



In the present paper we consider classical charged particles (electrons) in Weyl semimetal, with two Dirac cones of different chirality and non-zero Berry curvature, subject to the external electromagnetic Maxwell field. Due to the coupling between both $U(1)$ fields the equations of motion are further modified by the Berry curvature. As a result the kinetic theory is changed in a very nontrivial way. These modifications influence all transport characteristic of the system including chiral magnetic effect, anomalous Hall conductivity and other transport characteristics. The expected imprints of 
{\it dark matter} in the kinetic coefficients are proportional to the coupling $\alpha$.
Contrary to the previous proposals of measuring the electromagnetic fields produced by dark matter induced oscillating surface currents \cite{iwazaki2020,iwazaki2021,kishimoto2022} we propose to measure transport currents in the bulk. In the concluding section we shall comment on the 
{\it dark matter} detection feasibility via observation of the chiral magnetic effect.

Our first and main aim of this work is thus to derive the kinetic theory for the chiral system with non-trivial band topology in the presence of {\it dark sector}. 
It is well known that the standard Boltzmann kinetic theory must be properly generalized to study systems with Berry curvature. Berry curvature acts in momentum space like a magnetic field in real space~\cite{haldane2004,xiao2010} and introduces important changes in the kinetic equation. In order to treat the problem effectively we perform the transformation of the gauge fields to get rid of the {\it kinetic mixing } term.  To reach the goal we generalize the approach of the paper \cite{ste12} and derive the Boltzmann kinetic equation valid for Weyl semimetal subject to visible and {\it dark} fields.

\subsection{Background theory remarks}

To set the stage we now briefly review the field theoretical derivation of the kinetic theory~\cite{ste12} to describe Weyl quasiparticles, which are low energy excitations in numerous condensed matter compounds. The low energy Hamiltonian of the Weyl spin 1/2 particle with chirality $\lambda=\pm 1$ in condensed matter can be written as
\be
H_\lambda=\lambda ~\hbar~ v_F \mathbf{k}\cdot\boldsymbol{\sigma},
\label{weylh}
\ee
where $\boldsymbol{\sigma}$ is a vector of Pauli  $2\times 2$ matrices, $\mathbf{k}$ is the wave vector, and $v_F$ the Fermi velocity
We start with the phase space action~\cite{ste12} for a   particle with $\lambda=1$ and relativistic spectrum 
\be
\epsilon_\la(\mathbf{p})= \la v_F|\mathbf{p}|
\label{spec}
\ee
obtained by diagonalizing the relation (\ref{weylh}), where $\mathbf{p}$ is a momentum vector.  Fermi velocity  $v_F$  in condensed matter replaces the light velocity. For simplicity we assume $v_F=1$ and work in units with $\hbar=1$, $c=1$ and Boltzmann constant  $k_B=1$. Consequently the action reads 
\be
I=\int{(\mathbf{p}\cdot\mathbf{\dot{x}}-|\mathbf{p}|-\mathbf{a_p}\cdot\mathbf{\dot{p})}}dt.
\label{onefieldaction}
\ee
In the last equation $\mathbf{a_p}$ denotes the Berry connection related to the gauge independent Berry curvature vector $\boldsymbol{\Omega}$ by
\be
\boldsymbol{\Omega}=\mathbf{\nabla_p}\times \mathbf{a_p},
\ee
which for the system described by the equation (\ref{weylh}), can be found as~\cite{bernevig2013}
\be
\boldsymbol{\Omega}=\frac{\mathbf{p}}{2|\mathbf{p}|^3}.
\label{berry1}
\ee
It contains singularity for $\mathbf{p}=0$. Using minimal coupling to the electromagnetic scalar $\phi$ and vector $\mathbf{A}$ potentials one derives from (\ref{onefieldaction}), the equations of motions 
\ben
\label{eom1}
\mathbf{\dot{x}}=\mathbf{v_p}+\mathbf{\dot{p}}\times \boldsymbol{\Omega} \\
\mathbf{\dot{p}}=\mathbf{E}+\mathbf{\dot{x}}\times\mathbf{B},
\label{eom2}
\een
where $\mathbf{E,~B}$ denote electric and magnetic Maxwell fields, respectively. 
The solution to the classic Boltzmann kinetic equation for the distribution function is
modified by the modification of the velocity through the Berry curvature, which acts as a fictitious magnetic field in momentum space. This 'field' affects the transport characteristics of topological systems with chiral anomaly leading {\it inter alia} to anomalous Hall
~\cite{haldane2004} and chiral magnetic effects~\cite{fukushima2008}. 

The kinetic theory equations describe the time evolution in the phase space of
distribution function $f_\la(t,~x_\mu,~p_\mu)$, bounded with the right-handed  fermions for which $\la =1$ and left-handed ones with $\la = -1$. Total change of the distribution function in phase space results from the scattering processes

\be
\frac{d f_\la}{d t} =  \frac{\p f_\la}{\p t} + \frac{\p f_\la}{\p x_\mu} {\dot x}_\mu + \frac{\p f_\la}{\p p_\mu} {\dot p}_\mu = C[f_\la],
\label{dist}
\ee
where $C[f_\la]$ stands for the adequate collision integral. The specific functional form of the collision integral will be  discussed later on.

The modifications of the standard equations of motion given by (\ref{eom1}) and  (\ref{eom2}), affect the transport characteristics of systems with nontrivial topology of their spectrum. The mentioned existence of anomalous Hall effect
in a time reversal symmetry breaking materials is an important resulting feature. Chiral magnetic effect, which is the appearance of the current along the applied magnetic field is an important phenomenon first proposed in the context of particle physics to account for some effects in heavy-ion collisions~\cite{fukushima2008}. As stated earlier, our first aim is to search for the signatures of 
{\it dark sector} in these and other transport characteristics of Weyl semimetals.  Thus we shall first generalize the chiral kinetic theory to take the effect of dark field into account. 
Secondly we shall consider the  Weyl semimetal and calculate the dependence of the 
anomalous Hall and chiral magnetic effects
on the parameters of the model. Our study indicates that these transport parameters induced in Weyl semimetals by {\it dark sector} electric or magnetic fields show the same functional dependence with the amplitude scaled by the coupling between both $U(1)$ gauge fields.


The organization of the paper is as follows. In the next Sec. \ref{sec:darkphoton} we shall present the model  of {\it dark photon}
and find the generalized four-momentum of massive particle moving in two $U(1)$-gauge fields.
 The chiral kinetic theory is 
elaborated in Sec. \ref{sec:chiral}, where we have restricted our attention to the case of semiclassical approach and weak magnetic fields. 
Sec. IV is devoted to the collision-less case, while in Sec. V we consider collision case and pay attention to {\it dark photon} influence on magneto-transport properties.
In Sec. VI we pay attention to the effect of scattering and its influence on number/charge currents and conductivities. One also considers the case when the {\it dark matter} charge
is equal to zero, but the impact of {\it dark sector} survives by means of $\alpha$-coupling constant and {\it dark sector} magnetic field. Sec. VII concludes our investigations.

\section{The Model of dark photon} \label{sec:darkphoton}
In this section we shortly describe the {\it dark photon} model and propose a transformation of underlying gauge fields in order to simplify 
the underlying action, i.e., dispose of {\it kinetic mixing} term. In the process of this we obtain new gauge fields, being the mixture of the starting ones, with adequate factors
comprising $\alpha$-coupling constant. Next, one derives the equation of motion for charged particle affected by two gauge fields.

The action describing two coupled, massless gauge field is given by
\be
S_{M-dark~ photon} = \int  d^4x  \Big(
- F_{\mu \nu} F^{\mu \nu} - B_{\mu \nu} B^{\mu \nu} - {\alpha}F_{\mu \nu} B^{\mu \nu}
\Big),
\label{ac dm}
\ee  
where $\alpha$ is taken as a coupling constant. {\color{red} The model of massive dark photon will be briefly discussed in Section \ref{sec:massive}.}

 To get rid of the {\it kinetic mixing } term we define new gauge fields. They can be written  as 
\ben \label{transA}
\tA_\mu &=& \frac{\sqrt{2 -\alpha}}{2} \Big( A_\mu - B_\mu \Big),\\ \label{transB}
\tB_\mu &=& \frac{\sqrt{2 + \alpha}}{2} \Big( A_\mu + B_\mu \Big).
\een
As a result one leaves with only modified gauge fields, i.e., 
\be
 F_{\mu \nu} F^{\mu \nu} +
B_{\mu \nu} B^{\mu \nu} +  \alpha F_{\mu \nu} B^{\mu \nu}
\Longrightarrow
 \tF_{\mu \nu} \tF^{\mu \nu} +
\tB_{\mu \nu} \tB^{\mu \nu},
\ee
where we set $\tF_{\mu \nu} = \p_\mu \tA_{\nu}  - \p_\nu \tA_\mu$, 
and respectively $\tB_{\mu \nu} = \p_\mu \tB_{\nu}  - \p_\nu \tB_\mu$.
Just the action can be rewritten as
\be
S_{m} = \int d^4x  \Big(
- \tF_{\mu \nu} \tF^{\mu \nu} - \tB_{\mu \nu} \tB^{\mu \nu}
\Big).
\label{vdc}
\ee
Variation of the action (\ref{vdc}) with respect to $g_{\mu\nu},~\tA_\mu$ and $\tB_\mu$ reveals the following equations of motion for Maxwell {\it dark matter} system:
\be
\na_{\mu} \tF^{\mu \nu } = 0, \qquad \na_{\mu} \tB^{\mu \nu } = 0.
\label{fb}
\ee

 Having in mind relations (\ref{vdc}) and (\ref{fb}), we shall search for the equation of motion of charged particle in the background of the modified gauge fields. 
 However, the consistency of the approach requires the appropriate redefinitions of the charges coupled to original fields.
 The  resulting action of massive charged particle influenced by both visible and {\it dark matter} sectors is assumed to be 
\be
S = - \int m \sqrt{- ds^2} + \te_A \int \tA_{\mu} dx^\mu + \te_B \int \tB_\mu dx^\mu,
\ee
where  for transformed charges are defined as
\ben \label{cA}
\te_A &=& \frac{\sqrt{2 -\alpha}}{2} \Big( e - e_d \Big),\\ \label{cB}
\te_B &=& \frac{\sqrt{2 + \alpha}}{2} \Big( e + e_d \Big).
\een
In the above equations $e$ stands for the Maxwell charge, while $e_d$ is connected with {\it dark sector} one. 
The standard calculation leads to the following equation of motion
\be
m~\frac{D u^\mu}{d \tau} = \Big( \te_A \tF^{\mu \nu} +  \te_B \tB^{\mu \nu} \Big) u_\nu,
\ee
where $\frac{D}{d \tau}$ denotes in general case the covariant derivative with respect to the proper time.

As a result the four-momentum of the massive particle subject to two gauge fields 
may be written as
\be
P_\mu = m u_\mu +\te_A \tA_\mu +  \te_B  \tB_\mu.
\ee
We can notice that transforming the charges and fields back one arrives at 
\be
P_\mu = m u_\mu +   e A_\mu + e_d B_\mu  + \frac{\alpha}{2} \Big( e_d A_\mu + e B_\mu \Big).
\label{cons}
\ee
Having in mind the action (\ref{ac dm}), 
the above equation (\ref{cons}) provides the consistency in choosing transformed charges (\ref{cA})-(\ref{cB}) and fields (\ref{transA}) and (\ref{transB}).

It can be remarked that for the special case, when $e_d=0$, the above relation reduces to the following:
\be
P_\mu = m u_\mu +e(A_\mu + \frac{\alpha}{2}B_\mu).
\ee
This shows the modifications of the kinetic momentum by {\it dark sector} gauge field.

\section{Dark photon chiral kinetic theory} \label{sec:chiral}
 This section will be concerned with the derivation of basic relations for the kinetic chiral theory of {\it dark photon} case.
The theory in question describes the motion of particles in the regime where collisions are infrequent enough. This requires
\cite{ste15, wan91}
that $T \leq \sqrt{B^{(i)}} \leq \mu^{(i)}_\la$, where $i = \tF,~ \tB$, and $B^{(i)}$ is the magnetic field connected 
with the adequate gauge field, while $\mu^{(i)}_\la$ denotes the suitable chemical potential. In the aforementioned regime Landau quantization does not take place.

In order to derive the equations of motion for charge particle under the influence of external Maxwell and {\it dark matter} gauge fields, as well as, modified by the Berry flux, we generalize the action (\ref{onefieldaction}). 
Namely, we allow for the couplings of both effective charges $\te_A$ and $\te_B$ with the adequate effective gauge fields, in analogy to standard approach with one 
$U(1)$-gauge field \cite{ste12}
 \be
 I = \int_{t_1}^{t_2} dt \Bigg[ p_m {\dot x}^m+ \te_A~ \tA_m {\dot x}^m -  \te_A~ \Phi_{(\tF)} + \te_B~  \tB_m{\dot x}^m
 -  \te_B~ \Phi_{(\tB)} - \mid \vec p \mid - \la~a_{(p)}^m ~{\dot p}_m \Bigg],
 \label{action legendre}
 \ee
 and perform its variation to find the equations of motion generalising those given by equations (\ref{eom1}) and (\ref{eom2}).   It reveals that the equations of motion imply the following relations:
\ben
\label{pdot}
{\dot p}_a &=& \te_A \Big(
E_a^{(\tF)} + \ep_{abc}~ {\dot x}^b B^{c (\tF)} \Big) + \te_B \Big(
E_a^{(\tB)} + \ep_{abc} ~{\dot x}^b  B^{c (\tB)} \Big), \\
{\dot x}_a &=& {\hat p}_a + \ep_{abc} ~{\dot p}^b ~\Omega^c_\la,
\label{xdot}
\een
where one sets
\be
E_a^{(\tF)} = - \tF_{0a}, \qquad E_a^{(\tB)} = - \tB_{0a},
\ee
and
\be
B_a^{(\tF)} = \frac{1}{2} \ep_{abc} \tF^{bc}, \qquad B_a^{(\tB)} = \frac{1}{2} \ep_{abc} \tB^{bc}.
\ee

With the spectrum (\ref{spec}) and in our units assuming the Fermi velocity $v_F=1$, the particle velocity $v^i (p)=\frac{\partial \epsilon(p)}{\partial p_i}$ reduces to the direction  of the momentum vector
\be 
{\hat p}^a = \frac{p^a}{\mid \vec p \mid}.
\ee
We have denoted the Berry curvature vector of the Dirac cone $\lambda$ as $\boldsymbol{\Omega}_\la$. For the Hamiltonian (\ref{weylh}) it is given by formula (\ref{berry1}) and in new notation reads
\be
{\Omega}^a_\la =\la \frac{{\hat p}^a }{2 \mid \vec p \mid^2}.
\label{berry-a}
\ee

The solutions of the above relations (\ref{pdot}) and (\ref{xdot}) for ${\dot x}_\mu$ and ${\dot p}_\mu$ yield
\be
{K}_\la ~{\dot x}_a = 
{\hat p}_a  + \te_A ~\ep_{abc}~ E^{b (\tF)} {\Omega}^c_\la+  \te_B~\ep_{abc}~ E^{b (\tB)} \Omega^c_\la
+   {\hat p}_m \Omega^m_\la \Big(  \te_A ~B_a^{(\tF)} +  \te_B~B_a^{(\tB)} \Big),
\label{xprim}
\ee
and
\ben \label{pprim}
{K}_\la ~{\dot p}_a &=& \te_A ~E_a^{(\tF)} +  \te_B ~E_a^{(\tB)} 
+  \te_A~ \ep_{abc} ~{\hat p}^b B^{c (\tF)} + 
\te_B ~\ep_{abc} ~{\hat p}^b B^{c (\tB)} \\ \nonumber 
&+& \Omega_{a{}\la} \Big[ \te^2_A ~E_c^{(\tF)} B^{c (\tF)}  +  \te^2_B~E_c^{(\tB)} B^{c (\tB)}  +  \te_A \te_B~ \Big( 
E_c^{(\tB)} B^{c (\tF)}  + E_c^{(\tF)} B^{c (\tB)}  \Big) \Big],
\een
where the determinant of the matrix coefficients is given by
\be
K_\la = 1 + \te_A~B_c^{(\tF)} \Omega^c_\la +  \te_B~B_c^{(\tB)} \Omega^c_\la.
\label{klambda}
\ee
Substituting the  achieved results to the relation (\ref{dist}), we obtain all the desired ingredients for the chiral kinetic  equation, for the {\it dark photon} - Maxwell system.

Formulae (\ref{xprim}) - (\ref{klambda}) comprise the general equations of motion valid in chiral systems, with Berry curvature $\Omega_\lambda$ affected by two gauge fields.  
They generalize the standard equations of the chiral kinetic Boltzmann theory and constitute the main result of this section.
For instance it can be observed that assuming dark charge $e_d=0$, one finds that
\be
K_\la = 1 + e\Big(B_c^{(F)}+\frac{\alpha}{2}B_c^{(B)} \Big) \Omega^c_\la,
\label{klam2}
\ee
which clearly shows modification of the effective magnetic field due to to the coupling between visible and {\it dark sectors}.

From the set of equations it is visible that even without dark charge there exist modifications of the chiral  kinetic theory.
The adequate terms are all proportional to $\alpha$-coupling constant representing small corrections to the Maxwell fields. However, their presence modifies 
all measurable characteristics of the chiral materials and can be detected in specially designed experiments. 
In the typical setup, with external standard electric and magnetic  fields, these corrections are of order of $\alpha$ on top of values of order one and will be probably hard to detect. 
However, in the absence of external ${E}_\mu^{(F)}$ and $B_\mu^{(F)}$ electromagnetic fields, the  base signal is zero and the detection of probably small but non-zero signal induced by 
{\it dark magnetic} field ${B}_\mu^{(B)}$  can be possible and it may serve as an indication of the {\it dark sector} appearance. This scenario has similarities to the on-going experiments.

To see directly the expected  modifications of the transport characteristics of chiral system we shall derive explicit formulae for anomalous Hall effect (AHE) 
and chiral magnetic effect (CME) in the next section.

\section{Anomalous Hall and chiral magnetic effects: collision-less case}
Now we proceed to calculation of the anomalous Hall and chiral magnetic effects in the collision-less limit in which one supposes that $C[f_\la] = 0$. One assumes
 that the resulting solution of the homogeneous Boltzmann equation reduces to the Fermi - Dirac distribution function of the form as
\be
f^{(0)}_{\lambda}(\vec p)=\frac{1}{e^{\frac{{\epsilon}_{\lambda}(\vec p)-\mu}{k_BT}}+1},
\label{f-d-0}
\ee
where $T$ denotes temperature, $\mu$ chemical potential and $k_B$ is Boltzmann constant. 
As was mentioned in subsection A, in the Introduction, we shall use units where $\hbar = c = k_B =1$. We explicitly assume that
the spectrum depends on the handedness $\lambda$ of Weyl particles. 
 
It turns out that the presence of Berry curvature modifies the volume of phase-space \cite{xiao2010}.  As was revealed in \cite{ste12} this fact leads to
 the relation for the measure ${K_\la}$, which implies the following modified Liouville equation:
\ben \label{liu}
&{}& \frac{\p {K_\la}}{\p t} + \frac{\p}{\p x_m} \Big( {K_\la} ~{\dot x}_m \Big) + \frac{\p}{\p p_m} \Big( {K_\la} ~{\dot p}_m \Big) = \\ \nonumber
&=& 2\la  \pi ~\delta^3 (\vec p) ~ \Big[
\te^2_A~E_c^{(\tF)} B^{c (\tF)}  +  \te^2_B~ E_c^{(\tB)} B^{c (\tB)}  +
\te_A \te_B~\Big( E_c^{(\tB)} B^{c (\tF)}  + E_c^{(\tF)} B^{c (\tB)}  \Big) \Big].
\een
The right-hand side of the equation (\ref{liu}) was achieved by using the fact of the invariance of the phase space measure ${K_\la} d^3 {\vec x}~ d^3 {\vec p}/ (2 \pi)^3$ and the 
adequate components of the gauge field equations of motion and the relation for the Berry
monopole
\be
{\vec \na_{ p}} \boldsymbol{\Omega}_\la = 2 \pi \la  \delta^3 (\vec p).
\ee

Further, we define the number current density in the form 
\be
j^m _\la= \int \frac{d^3 p}{(2 \pi)^3} {K_\la} f_\la~{\dot x}^m.
\label{curr-gen}
\ee
Consequently, taking into account the relations (\ref{xprim}) and (\ref{f-d-0}), we obtain the total number current in the collision-less limit 
\be
j^a_\la = j_{(o) \la }^a + j_{(Hall~ an) \la}^{a  (\tF)} + j_{(Hall ~an) \la}^{a (\tB)}+ j_{(CME) \la}^{a  (\tF)} + j_{(CME) \la}^{a (\tB)},
\label{current}
\ee
where the first number current in the above expression
\be  
 j_{(o) \la}^a = \int \frac{d^3 p}{(2 \pi)^3} f^{(0)}_\la~ {\hat p}^a,
 \ee
does not depend on external fields and vanishes as the equilibrium property of the system. Other terms in the equation (\ref{current}) denote 
currents, linear in electric and magnetic fields related to both $U(1)$ sectors. Namely, they constitute the anomalous Hall number currents 
 \be
 j_{(Hall~ an) \la}^{a (\tF)} = \te_A \int \frac{d^3 p}{(2 \pi)^3} f^{(0)}_\la  ~\ep^{abc}~ E_{b}^{ (\tF)}  \Omega_{c \la}, \qquad
 j_{(Hall ~an)\la}^{a (\tB)} = \te_B\int \frac{d^3 p}{(2 \pi)^3} f^{(0)}_\la  ~\ep^{abc}~ E_{b}^{ (\tB)} \Omega_{c \la},
\ee
and the chiral magnetic number currents (along the direction of the magnetic fields)
\be
 j_{(CME)\la}^{a (\tF)} =  \te_A \int \frac{d^3 p}{(2 \pi)^3} f^{(0)}_\la ~ {\hat p}_m \Omega^m _\la ~B^{a (\tF)},
 \qquad   j_{(CME) \la}^{a (\tB)} = \te_B\int \frac{d^3 p}{(2 \pi)^3} f^{(0)}_\la ~ {\hat p}_m \Omega^m_\la B^{a (\tB)}.
\label{jcme}
\ee
Similarly as  in Ref. \cite{ste12}, 
one uses the relation $E = \mid \vec p \mid$, and overbar to indicate
 the averaging over the unit sphere of directions of vector $\hat p_i$, we arrive at the expression
\be
j^{a (\tF), (\tB)}_{(CME) \la} = \te_{A (B)}~\frac{B^{a (\tF)} (B^{a{(\tB)}})}
{4 \pi^2} \int_0^\infty dE~\overline{  f^{(0)}_\la(E, \hat p)}.
\ee
To get the last formulae we introduced expression (\ref{berry-a}) for Berry curvature into (\ref{jcme}), changed variables using $E=|p|$, assumed that in general $f(\vec{p})=f(E,\hat{p})$  and denoted $\overline{  f^{(0)}_\la(E, \hat p)}$ by
\be
\overline{  f^{(0)}_\la(E, \hat p)}=\frac{1}{4\pi}\int_{S^2_{\hat p}} d{\bf {\hat p}}~f^{(0)}_\la(E, \hat p).
\ee

Moreover when one assumes the isotropic Fermi-Dirac distribution (i.e., for $\overline{  f^{(0)}_\la(E, \hat p)}=f^{(0)}(E)$), the  equilibrium currents along the directions of the magnetic fields are induced 
by both gauge fields describing Maxwell and {\it dark matter} sectors  and require finite and different chemical potential $\mu$ of chiral fermions. Namely the number currents for each chirality are given by  
\be
j^{a(\tF), (\tB)}_{(CME)\la} = \te_{A(B)}~\la~\frac{\mu_\la}{4 \pi^2} ~B^{a (\tF)}( B^{a (\tB)}).
\label{no-curr-ahe}
\ee
The charge currents densities are calculated in the next section.

\subsection{Charge current densities} \label{sec:charge-curr}
Let us remark that in order to obtain charge current densities and connected transport characteristics, one ought to  multiply the quantities achieved previously as in 
the equation (\ref{no-curr-ahe})  by corresponding charges given by the relations (\ref{cA}) and (\ref{cB}). Namely they are provided by 
\be
\tj_{(Hall~ an) \la}^{a (\tF)} = \te_A~ j_{(Hall~ an) \la}^{a (\tF)} , \qquad
 \tj_{(CME)\la}^{a (\tF)}  = \te_A ~j_{(CME)\la}^{a (\tF)},
\ee
for the field $\tF$ and corresponding effective charges $\te_A$ and respectively the analogous forms for $\tB$ field multiplied by $\te_B$. 
The conductivities connected with adequate current densities are defined in a standard way.

For the case when $e_d=0$, the total charge current along the applied magnetic field (known as chiral magnetic effect) in the node $\lambda$ reads
\be
\tj^a_{(CME)\lambda}=\tj_{(CME)\la}^{a (\tF)} +  \tj_{(CME) \la}^{a (\tB)} = 
 e^2 \int \frac{d^3 p}{(2 \pi)^3} f^{(0)}_\la ~ {\hat p}_m \Omega^m _\la ~\Big( \beta_F B^{a (F)} +\beta_B B^{a (B)} \Big),
 \ee
and for the isotropic Fermi-Dirac distribution, one gets
\be
\tj^a_{(CME)\lambda}=e^2\la\frac{\mu_\la}{2\pi^2} \Big( \beta_F B^{a (F)} +\beta_B B^{a (B)} \Big),
\ee
where we have denoted
\be
\beta_F=\left(\frac{2+\alpha}{4}\right)^{\frac{3}{2}}+\left(\frac{2-\alpha}{4}\right)^{\frac{3}{2}},
\ee
\be
\beta_B=\left(\frac{2+\alpha}{4}\right)^{\frac{3}{2}}-\left(\frac{2-\alpha}{4}\right)^{\frac{3}{2}}.
\ee
In a similar manner one finds that
\be
\tj_{(Hall~ an) \la}^{a}= \tj_{(Hall~ an) \la}^{a (\tF)} + \tj_{(Hall ~an)\la}^{a (\tB)} =
e^2\int \frac{d^3 p}{(2 \pi)^3} f^{(0)}_\la  ~\ep^{abc} \Big( \beta_F E_b^{(F)}
+\beta_B E_b^{(B)} \Big)\Omega_{c \la}.
\ee

\section{Chiral anomaly in the presence of two gauge fields $\tF_{\mu \nu}$ and $\tB_{\mu \nu}$ }

The presence of two gauge fields affects the value of the anomaly. To get some information on it in the theory in question we proceed as in~\cite{ste12}, by first defining
phase space currents $(Kf^{(0)},Kf^{(0)}\mathbf{\dot{x}},Kf^{(0)}\mathbf{\dot{p}})$, formulate equation similar to (\ref{liu}) and integrate it over all momenta.
 Consequently we arrive at the following expression:
\be 
\frac{\p n_\la}{\p t} + \na_a j^a_\la = \frac{\la}{4 \pi^2} f^{(0)}_{\la}(0) \Big[
\te^2_A~E_c^{(\tF)} B^{c (\tF)}  + \te^2_B~ E_c^{(\tB)} B^{c (\tB)}  
+ \te_A \te_B \Big(
E_c^{(\tB)} B^{c (\tF)}  +  E_c^{(\tF)} B^{c (\tB)}  \Big)\Big],
\label{chir}
\ee
where the particle number is given by
\be
n_\la = \int \frac{d^3 p}{(2 \pi)^3} ~{K_\la}~f_\la,
\label{num}
\ee
while $f^{(0)}_{\la}(0)$ is the value of the distribution function $f^{(0)}_\la$ for $p_i$ equal to zero. For the Fermi-Dirac distribution at zero temperature 
and for the non-zero (positive) values of the chemical potentials, $f^{(0)}_{\la}(0)$ is equal to one. 
Setting to zero all the components of {\it dark matter} sector field and $\alpha$-coupling constant, one obtains the limit of electromagnetic anomaly~\cite{ste12}.

As was remarked in Ref. \cite{ste12}, the above calculations should be taken with a great care, because of the fact that we integrated over the whole phase space, including the
singular point where $\vec p = 0$.
At this point the classical description is not valid and it is proposed that the region around the singularity ought to be excluded. However the 
classical description is reliable for $\vec p$ outside the infinitesimal region near singular point in the phase space manifold.

The relation (\ref{chir}) implies that the particle number described by (\ref{num}) around the Fermi surface 
 is not conserved for single chirality, due to the non-zero
value of the right-hand side of the relation. The non-conservation is caused by the existence of electric and magnetic fields connected with  the considered
gauge fields $\tF_{\mu \nu}$ and $\tB_{\mu \nu}$. Moreover prior to the ordinary Maxwell case we obtained that the mixture composed of electric and magnetic
fields pertaining to the different gauge sectors, also plays the role the role in the non-conservation.
This fact constitutes the additional new effect emerging form taking into account the {\it dark matter} sector coupled to the Maxwell electrodynamics via {\it kinetic
mixing} term.
 
 The conservation of the total charge requires that the total current is also conserved. This is indeed the case as the right hand side of (\ref{chir}) 
 vanishes, when summed up over two Weyl nodes. However, the chiral current is not conserved and subtracting  both sides of equation (\ref{chir}) for both 
 signs of the chirality $\lambda$ gives the chiral anomaly, which is twice the 
 right-hand side of the above equation. We conclude with the remark that both gauge fields contribute to chiral anomaly.

\section{Effect of scattering}
In previous section we have considered collision-less limit. 
In what follows we shall consider the collision term in the relaxation time approximation with a single relaxation time corresponding to intra node scatterings.
The collision integral is given by 
\be
C_1[f_\la] = - \frac{\delta f_\la}{\tau},
\ee
where $\delta f_\la = f_\la-f^{(0)}_{\la}$, $f^{(0)}_{ \la}$ is the Fermi-Dirac distribution function. This term constitutes modification on the right-hand side of the continuity relation (\ref{chir}). 

As our aim is to consider the homogeneous and stationary case, 
the linearized form of the Boltzmann equation is easy to solve for the correction to the equilibrium distribution function.  The obtained solution 
reveals that one can split out-of-equilibrium distribution function $\delta f_\la $ as 
\be
\delta\ f_\la = \delta f^{(O)}_\la + \delta f^{(Hall ~an)}_\la + \delta f^{(CME)}_\la,
\label{deltaf-relax}
\ee
where $\delta f^{(O)}_\la$ stand for ordinary or Ohmic part of the correction and we have denoted the above ingredients by
\ben 
\delta f^{(O)}_\la &=& - \frac{\tau}{{K_\la}} \frac{\p f_{0 \la}}{\p p^i} \Big[  \te_A~ E^{i (\tF)} +  \te_B~ E^{i (\tB)} \Big],\\
\delta f^{(Hall~an)}_\la &=& - \frac{\tau}{{K_\la}} \frac{\p f_{0 \la}}{\p p^i} \Big[ \te_A~ \ep^{ibc} {\hat p}_b B_c^{(\tF)} +  \te_B~\ep^{ibc} {\hat p}_b B_c^{(\tB)} \Big],\\
\delta f^{(CME)}_\la &=& - \frac{\tau}{{K_\la}} \frac{\p f_{0 \la}}{\p p^i} \Big[ \te^2_A ~E_c^{(\tF)} B^{c (\tF)}  + \te^2_B ~E_c^{(\tB)} B^{c (\tB)}  \\ \nonumber
&+& 
 \te_A \te_B~ (
E_c^{(\tB)} B^{c (\tF)}  + E_c^{(\tF)} B^{c (\tB)} )  \Big] \Omega^i_\la.
\een
Consequently the number current operators can be decomposed as (see the equation (\ref{current}) in the latter section)
\ben \label{scat-cur-O}
j_{(O) \la}^m &=& \int \frac{d^3 p}{(2 \pi)^3} ~{\hat p}^m~\delta f_\la,\\
\label{scat-cur-ahe}
j_{(Hall~an) \la}^m &=&  \int \frac{d^3 p}{(2 \pi)^3}~\ep^{mbc} ~ \Omega_{c \la} ~\Big( \te_A~ E_b^{(\tF)} +  \te_B~ E_b^{(\tB)} \Big)~\delta f_\la,\\
\label{scat-cur-cme}
j_{(CME) \la}^m &=&\int \frac{d^3 p}{(2 \pi)^3} ~{\hat p}_c \Omega^c_\la ~\Big( \te_A~ B^{m (\tF)} +  \te_B~ B^{m (\tB)} \Big)~\delta f_\la.
\een
The proportionality of the corrections to the scattering time $\tau$ clearly shows that the above are the corrections due to the collisions
 in the system. In the following we shall analyze the contributions of visible and {\it dark matter} sectors, to various parts of the current.

\subsection{Conductivities for $\tF_{\mu \nu}$ and $\tB_{\mu \nu}$ field strengths}
This subsection will be devoted to the magneto-transport properties influenced by {\it dark matter } sector.  Because  the anomalous Hall current is always transverse to the electric fields $E_a^{(\tF)}$ and $E_a^{(\tB)}$, so it will not be amiss to ignore it \cite{dan18}.

In the theory under consideration we shall consider conductivities associated with $\tF_{\mu \nu}$ and $\tB_{\mu \nu}$ field strengths, respectively.
The total conductivity bounded with $\tF_{\mu \nu}$ and the handedness $\la$, can be written as the derivative of the number current with respect to $E^{b (\tF)}$ multiplied by the corresponding charge ($\te_A$). It yields
\be
\sigma_b{}^a (\tF, \la) = \te_A \frac{\p}{\p E^{b (\tF)}} \Big( j_{(O) \la }^a + j_{(CME) \la}^a \Big) = \sigma_b{}^{a (1)} (\tF) + \sigma_b{}^{a (2)} (\tF) + \sigma_b{}^{a (3)} (\tF),
\ee
where one denotes
\be
\sigma_b{}^{a (1)} (\tF) = \int \frac{d^3 p}{(2 \pi)^3} ~{\hat p}^a~\frac{\tau}{{K_\la}}~ \te_A^2~\Big(- \frac{\p f_{0 \la}}{\p p^b} \Big),
\ee
and $\sigma_b{}^{a (2)} (\tF)$ implies the following:
\ben
\sigma_b{}^{a (2)} (\tF) &=& \int \frac{d^3 p}{(2 \pi)^3} ~{\hat p}^a~\frac{\tau}{{K_\la}}\Big(- \frac{\p f_{0 \la}}{\p p^i} \Big) \Omega^i_\la ~ \te_A^2
\Big( \te_A~B_b^{(\tF)} + \te_B~ B_b^{(\tB)} \Big) \\
&+& \int \frac{d^3 p}{(2 \pi)^3} ~{\hat p}_c \Omega^c_\la~\frac{\tau}{{K_\la}} \Big(- \frac{\p f_{0 \la}}{\p p^b} \Big) ~\te_A^2
\Big( \te_A~B^{a (\tF)} +  \te_B~B^{a (\tB)} \Big),
\een
while for $\sigma_b{}^{a (3)} (\tF)$ one has the relation
\be
 \sigma_b{}^{a (3)} (\tF) = \int \frac{d^3 p}{(2 \pi)^3} ~{\hat p}_c \Omega^c_\la~\frac{\tau}{{K_\la}}\Big(- \frac{\p f_{0 \la}}{\p p^i} \Big) \Omega^i_\la ~ \te_A^2
 \Big(  \te_A B^{a (\tF)} +  \te_B B^{a (\tB)} \Big)\Big(  \te_A B_b^{(\tF)} + \te_B  B_b^{(\tB)} \Big).
 \ee

On the  other hand, for $\tB_{\mu \nu}$ field strength and the adequate value of $\la$, the total conductivity connected with the number currents, is given by
\be
\sigma_b{}^a (\tB, \la) =\te_B \frac{\p}{\p E^{b (\tB)}} \Big( \tj_{(O) \la }^a + \tj_{(CME) \la}^a \Big) = \sigma_b{}^{a (1)} (\tB) + \sigma_b{}^{a (2)} (\tB) + \sigma_b{}^{a (3)} (\tB),
\ee
where we set
\be
\sigma_b{}^{a (1)} (\tB) = \int \frac{d^3 p}{(2 \pi)^3} ~{\hat p}^a~\frac{\tau}{{K_\la}} \te_B^2~\Big(- \frac{\p f_{0\la}}{\p p^b} \Big),
\ee
and denote 
\ben
\sigma_b{}^{a (2)} (\tB) &=& \int \frac{d^3 p}{(2 \pi)^3} ~{\hat p}^a~\frac{\tau}{{K_\la}}\Big(- \frac{\p f_{0 \la}}{\p p^i} \Big) \Omega^i _\la~ \te_B^2
\Big(  \te_A B_b^{(\tF)} + \te_B B_b^{(\tB)} \Big) \\
&+& \int \frac{d^3 p}{(2 \pi)^3} ~{\hat p}_c \Omega^c_\la~\frac{\tau}{{K_\la}} \Big(- \frac{\p f_{0 \la}}{\p p^b} \Big) ~ \te_B^2
\Big( \te_A B^{a (\tF)} +  \te_B B^{a (\tB)} \Big),
\een
while $\sigma_b{}^{a (3)} (\tB)$ can be written in the form as follows:
\be
 \sigma_b{}^{a (3)} (\tB) = \int \frac{d^3 p}{(2 \pi)^3} ~{\hat p}_c \Omega^c_\la~\frac{\tau}{{K_\la}}\Big(- \frac{\p f_{0 \la}}{\p p^i} \Big) \Omega^i_\la ~ \te_B^2
 \Big(  \te_A B^{a (\tF)} +  \te_B B^{a (\tB)} \Big)\Big( \te_A B_b^{(\tF)} +  \te_B B_b^{(\tB)} \Big).
 \ee
The obtained results are valid for an arbitrary choice of the components of gauge fields in question.

\subsection{Limiting case for $e_d=0$}\label{sec:ed0}

For completeness of the results we quote the relations for the adequate conductivities bounded with charge current densities, in the limit when {\it dark matter} charge is equal to zero.
They imply the following:
\ben \nonumber \label{ed1}
\sigma_b{}^{a (1)} (\tF) &=& \int \frac{d^3 p}{(2 \pi)^3} ~{\hat p}^a~\frac{\tau}{{K_\la}}~ \Big( \frac{\sqrt{2 -\alpha}}{2} \Big)^2 e^2
~\Big(- \frac{\p f_{0 \la}}{\p p^b} \Big),\\
\sigma_b{}^{a (2)} (\tF) &=& \int \frac{d^3 p}{(2 \pi)^3} ~{\hat p}^a~\frac{\tau}{{K_\la}}\Big(- \frac{\p f_{0 \la}}{\p p^i} \Big) \Omega^i_\la ~ e^3 \Big( \frac{\sqrt{2 -\alpha}}{2} \Big)^2
\Big( B_b^{(F)} + \frac{\alpha}{2} B_b^{(B)} \Big) \\ \nonumber
&+& \int \frac{d^3 p}{(2 \pi)^3} ~{\hat p}_c \Omega^c_\la~\frac{\tau}{{K_\la}} \Big(- \frac{\p f_{0 \la}}{\p p^b} \Big) ~e^3 \Big( \frac{\sqrt{2 -\alpha}}{2} \Big)^2
\Big( B_b^{(F)} + \frac{\alpha}{2} B_b^{(B)} \Big),\\ \nonumber
 \sigma_b{}^{a (3)} (\tF) &=& \int \frac{d^3 p}{(2 \pi)^3} ~{\hat p}_c \Omega^c_\la~\frac{\tau}{{K_\la}}\Big(- \frac{\p f_{0 \la}}{\p p^i} \Big) \Omega^i_\la ~  e^3 \Big( \frac{\sqrt{2 -\alpha}}{2} \Big)^2
 \Big( B^{a (F)} +  \frac{\alpha}{2}B^{a (B)} \Big)\Big(  B_b^{(F)} + \frac{\alpha}{2}B_b^{(B)} \Big),
 \een
and for $\tB_{\mu \nu}$ field strength, we obtain 
\ben \nonumber \label{ed2}
\sigma_b{}^{a (1)} (\tB) &=& \int \frac{d^3 p}{(2 \pi)^3} ~{\hat p}^a~\frac{\tau}{{K_\la}}~ \Big( \frac{\sqrt{2 +\alpha}}{2} \Big)^2 e^2
~\Big(- \frac{\p f_{0 \la}}{\p p^b} \Big),\\
\sigma_b{}^{a (2)} (\tB) &=& \int \frac{d^3 p}{(2 \pi)^3} ~{\hat p}^a~\frac{\tau}{{K_\la}}\Big(- \frac{\p f_{0 \la}}{\p p^i} \Big) \Omega^i_\la ~ e^3 \Big( \frac{\sqrt{2 +\alpha}}{2} \Big)^2
\Big( B_b^{(F)} + \frac{\alpha}{2} B_b^{(B)} \Big) \\ \nonumber
&+& \int \frac{d^3 p}{(2 \pi)^3} ~{\hat p}_c \Omega^c_\la~\frac{\tau}{{K_\la}} \Big(- \frac{\p f_{0 \la}}{\p p^b} \Big) ~e^3 \Big( \frac{\sqrt{2 + \alpha}}{2} \Big)^2
\Big( B_b^{(F)} + \frac{\alpha}{2} B_b^{(B)} \Big),\\ \nonumber
 \sigma_b{}^{a (3)} (\tB) &=& \int \frac{d^3 p}{(2 \pi)^3} ~{\hat p}_c \Omega^c_\la~\frac{\tau}{{K_\la}}\Big(- \frac{\p f_{0 \la}}{\p p^i} \Big) \Omega^i_\la ~  e^3 \Big( \frac{\sqrt{2 +\alpha}}{2} \Big)^2
 \Big( B^{a (F)} +  \frac{\alpha}{2}B^{a (B)} \Big)\Big(  B_b^{(F)} + \frac{\alpha}{2}B_b^{(B)} \Big),
 \een
where now $K_\la$ is given by the relation (\ref{klam2}).

From experimental point of view the  sums of  terms $i.e.$ $\sigma_b{}^{a (i)} (\tF) + \sigma_b{}^{a (i)} (\tB)$ matter and one gets the following expressions:
\ben \nonumber \label{ed3}
\sigma_b{}^{a (1)}  &=& \int \frac{d^3 p}{(2 \pi)^3} ~{\hat p}^a~\frac{\tau}{{K_\la}}~ e^2
~\Big(- \frac{\p f_{0 \la}}{\p p^b} \Big),\\
\label{cme-fin}
\sigma_b{}^{a (2)} &=& \int \frac{d^3 p}{(2 \pi)^3} ~{\hat p}^a~\frac{\tau}{{K_\la}}\Big(- \frac{\p f_{0 \la}}{\p p^i} \Big) \Omega^i_\la ~ e^3 
\Big( B_b^{(F)} + \frac{\alpha}{2} B_b^{(B)} \Big) \\ \nonumber
&+& \int \frac{d^3 p}{(2 \pi)^3} ~{\hat p}_c \Omega^c_\la~\frac{\tau}{{K_\la}} \Big(- \frac{\p f_{0 \la}}{\p p^b} \Big) ~e^3 
\Big( B_b^{(F)} + \frac{\alpha}{2} B_b^{(B)} \Big),\\ \nonumber
 \sigma_b{}^{a (3)} &=& \int \frac{d^3 p}{(2 \pi)^3} ~{\hat p}_c \Omega^c_\la~\frac{\tau}{{K_\la}}\Big(- \frac{\p f_{0 \la}}{\p p^i} \Big) \Omega^i_\la ~  e^3 
 \Big( B^{a (F)} +  \frac{\alpha}{2}B^{a (B)} \Big)\Big(  B_b^{(F)} + \frac{\alpha}{2}B_b^{(B)} \Big),
 \een
where now the $K_\lambda$ is given by the relation (\ref{klam2}).
It is the term $\sigma_b{}^{a (2)}$ which can be regarded as describing the 
magnitude of the {\it dark field} induced chiral magnetic conductivity.

\section{Dark photon massive case} \label{sec:massive}
To finish with, let us give some remarks concerning the {\it dark photon massive} case and derivation of equations of motion
for charge particle influenced by Maxwell and {\it massive dark matter}  sectors.

The action for the massive case of {\it dark photon} will be provided by
\be
S_{DM} = \int  d^4x  \Big(
- F_{\mu \nu} F^{\mu \nu} - B_{\mu \nu} B^{\mu \nu} - {\alpha}F_{\mu \nu} B^{\mu \nu} - \frac{m_{DM}^2}{2}  B_\mu B^\mu
\Big),
\label{mass dm}
\ee  
where as in the previous sections, $\alpha$ stands for a coupling constant.

 In the case under consideration, in order to eliminate the {\it kinetic mixing } term we define new gauge fields in the forms as follows:
\be
\tA_\mu = A_\mu + \frac{\alpha}{2}B_\mu,\qquad
\tB_\mu = \sqrt{1 - \frac{\alpha^2}{4}} B_\mu.
\label{mass case}
\ee
The above transformation enables us to rewrite the action (\ref{mass dm})
\be
 F_{\mu \nu} F^{\mu \nu} +
B_{\mu \nu} B^{\mu \nu} +  \alpha F_{\mu \nu} B^{\mu \nu} + \frac{m_{DM}^2}{2} B_\mu B^\mu
\Longrightarrow
 \tF_{\mu \nu} \tF^{\mu \nu} +
\tB_{\mu \nu} \tB^{\mu \nu} + \frac{{\tilde m}_{DM}^2}{2} \tB_\mu \tB^\mu. 
\ee
where we denoted by ${\tilde m}_{DM}$
\be
{\tilde m}_{DM}^2  = \frac{m_{DM}^2}{1 - \frac{\alpha^2}{4}}.
\ee
The equations of motion for $\tA_\mu$ and $\tB_\mu$ gauge fields now yield
\be
\na_\mu \tF^{\mu \nu} = 0, \qquad  \na_\mu \tB^{\mu \nu} - \frac{ {\tilde m}_{DM}^2 }{4} \tB^\nu =0.
\label{massive}
\ee
The equations of motion for charge particle under the influence of external Maxwell and massive {\it dark matter} gauge fields, will be derived from 
the action (\ref{action legendre}), where now $\tB_\mu$ field will obey the equation of motion (\ref{massive}) and
 the transformed charges imply
 \be
 \te_{A} = e + \frac{\alpha}{2} e_d, \qquad \te_B = \sqrt{1 - \frac{\alpha^2}{4}} e_d.
 \label{charges massive}
 \ee
 After performing variation, the equations of motion for ${\dot p}_a$ and ${\dot x}_a$, will be provided by the relations (\ref{pdot}) and (\ref{xdot}), with the new meaning
 of the gauge fields (\ref{mass case}) and charges (\ref{charges massive}).

In the context of the  results presented in subsection \ref{sec:ed0} for $e_d=0$ case, we give some remarks concerning the  estimations of the magnitude of the chiral magnetic effect induced by dark matter induced magnetic field. First of all, let us suppose that we are able to conduct an experiment in which one gets rid of the influence of magnetic Maxwell field, i.e., $B_a^{(F)} =0$.

The recent measurements of the local {\it dark matter} densities conducted by LAMOST DR5 and Gaia DR2 experiments \cite{guo20,loe22}, reveal that our Galactic disc is immersed in {\it dark matter} halo with a characteristic mass density $\rho_{DM}=0.5~ GeV/cm^3 $. 
On the basis of the action (\ref{mass dm}) the {\it dark matter} density is  expected to be bounded with the average {\it dark field}  of magnitude $B$ depending on {\it dark photon} mass \cite{beringer2012,iwazaki2020,iwazaki2021,kishimoto2022} 
\be
\rho_{DM} =m_{DM}^2 B_c^2/2.
\label{dark-dens}
\ee
Having in mind  equation for {\it hidden photon} field given in Ref. \cite{kishimoto2022}
\be
\vec{B}=\vec{B}_c\cos(m_{DM}t-m_{DM}\vec{v}_{DM}\cdot\vec{x}),
\label{vector-B}
\ee
one finds z-component of the {\it dark matter} magnetic induction 
\be
B^{(B)z}=m_{DM}(B_c^{y} v_{DM}^x-B_c^{x} v_{DM}^y)\sin(m_{DM}t-m_{DM}\vec{v}_{DM}\cdot\vec{x}).
\ee
Due to the fact that the direction of the {\it dark matter} $\vec{B}$ field is expected to be constant over a large regions in space,
 one finds that on average  the magnetic induction is of the order of $B^{(B)}_c=|\vec{v}_{DM}|m_{DM}B_c$. Using  equation (\ref{dark-dens}) for $\rho_{DM}$ we find $B^{(B)}_c=|\vec{v}_{DM}|\sqrt{2\rho_{DM}}$. Note, that in natural units both sides have mass dimension $eV^2$, as they should. Interestingly, 
the magnetic induction does not depend on the mass of {\it dark photon}. However, the frequency of the field depends on $m_{DM}$, as is seen from the relation (\ref{vector-B}). 

We take {\it dark matter} velocity as $|\vec{v}_{DM}|\approx 10^{-3}$ [24]. Thus $B^{(B)}_c \approx 2.77 \times 10^{-6} ~eV^2$ in natural units, or
expressed in Tesla the {\it dark matter} induced magnetic induction  $B^{(B)}_c \approx 4 \times 10^{-9}~ T$, which amounts to the tiny fraction of the Earth magnetic field. 

Thus we conclude that  even for $\alpha$ not much smaller than about $10^{-6}$ \cite{tho22}, the CME signal as given by Eq. (\ref{ed3}) which is proportional to $\alpha B^{(B)}_c $,
(by the direct inspection it can be checked that the same situation holds for {\it dark massive photon} case),
will be rather difficult to measure directly. 


\section{Conclusions}

In the paper we have generalized the chiral kinetic theory taking into account the effects of two coupled $U(1)$-gauge fields. One is the standard Maxwell field, while the other pertains to the {\it hidden sector}. Both fields are coupled by the so-called kinetic mixing term.

Our main aim was to find the modifications of the Boltzmann kinetic equation by {\it dark photon} and envisage the additional effects in transport coefficients caused by {\it dark sector}. The obtained results show that chiral Boltzmann theory is severally modified. The distribution function and the resulting currents are affected by both $U(1)$ fields.
The non-trivial topology of Dirac/Weyl semimetals leads to the chiral magnetic effect, which constitutes the current flow along the applied magnetic field. It happens that the  aforementioned effect may also result from the presence of {\it dark sector} induced magnetic field as is obvious from equation (\ref{cme-fin}).
Even though its direct detection in bulk experiments, with an applied external $E_\mu^{(F)}$ and $B_\mu^{(F)}$ Maxwell fields may be difficult, as it appears as an additive correction  proportional to $\alpha$, it could be measurable in the experiments without applied external magnetic field $B_\mu^{(F)}=0$. In such a case signal proportional to $\alpha B_\mu^{(B)}$ stemming from {\it dark magnetic} field becomes the only non-zero signal. The order of magnitude estimation demonstrate that its direct observation will be difficult, if feasible at all.

\acknowledgements
The authors would like to thank the unknown Referee for the useful comments on the first version of the paper.
M.R and K.I.W were partially supported by Grant No. 2022/45/B/ST2/00013 of the National Science Center, Poland.



\begin{thebibliography}{99}

%
\def\cmp#1#2#3#4{\emph{#4}, \emph{ Commun. Math. Phys.} {\bf #1} (#3) #2}
\def\lmp#1#2#3#4{\emph{#4}, \emph{ Lett. Math. Phys.} {\bf #1} (#3) #2}
\def\hpa#1#2#3#4{\emph{#4}, \emph{ Hell. Phys. Acta} {\bf #1} (#3) #2}
\def\grg#1#2#3#4{\emph{#4}, \emph{ Gen. Rel. Grav.} {\bf #1} (#3) #2}
\def\pr#1#2#3#4{\emph{#4}, \emph{ Phys. Rev.} {\bf #1} (#3) #2}
\def\prl#1#2#3#4{\emph{#4}, \emph{ Phys. Rev. Lett.} {\bf #1}, #2 (#3)}
\def\prd#1#2#3#4{\emph{#4}, \emph{ Phys. Rev. D} {\bf #1}, #2 (#3)}

\def\prb#1#2#3#4{\emph{#4}, \emph{ Phys. Rev. B} {\bf #1}, #2 (#3) }
\def\prx#1#2#3#4{\emph{#4}, \emph{ Phys. Rev. X} {\bf #1} (#3) #2}
\def\pl#1#2#3#4{\emph{#4}, \emph{ Phys. Lett.} {\bf #1} (#3) #2}
\def\pla#1#2#3#4{\emph{#4}, \emph{ Phys. Lett. A} {\bf #1} (#3) #2 }
\def\plb#1#2#3#4{\emph{#4}, \emph{ Phys. Lett. B} {\bf #1}, #2 (#3)}
\def\prep#1#2#3#4{\emph{#4}, \emph{ Phys. Reports} {\bf #1}, #2 (#3)}
\def\phys#1#2#3#4{\emph{#4}, \emph{ Physica} {\bf #1} (#3) #2}
\def\jcp#1#2#3#4{\emph{#4}, \emph{ J. Comput. Phys.} {\bf #1} (#3) #2}
\def\jmp#1#2#3#4{\emph{#4}, \emph{ J. Math. Phys.} {\bf #1} (#3) #2}
\def\jpm#1#2#3#4{\emph{#4}, \emph{ J. Phys. A: Math. Gen.} {\bf #1} (#3) #2}
\def\cpr#1#2#3#4{\emph{#4}, \emph{ Computer Phys. Rept.} {\bf #1} (#3) #2}
\def\cqg#1#2#3#4{\emph{#4}, \emph{ Class. Quant. Grav.} {\bf #1} (#3) #2}
\def\cma#1#2#3#4{\emph{#4}, \emph{ Computers Math. Applic.} {\bf #1} (#3) #2}
\def\mc#1#2#3#4{\emph{#4}, \emph{ Math. Compt.} {\bf #1} (#3) #2}
\def\apj#1#2#3#4{\emph{#4}, \emph{ Astrophys. J.} {\bf #1} (#3) #2}
\def\apjs#1#2#3#4{\emph{#4}, \emph{ Astrophys. J. Suppl.} {\bf #1} (#3) #2}
\def\apjl#1#2#3#4{\emph{#4}, \emph{ Astrophys. J. Lett.} {\bf #1} (#3) #2}
\def\acta#1#2#3#4{\emph{#4}, \emph{ Acta Astronomica} {\bf #1} (#3) #2}
\def\apl#1#2#3#4{\emph{#4}, \emph{ Ann. Physik. (Leipzig)} {\bf #1} (#3) #2}
\def\amjp#1#2#3#4{\emph{#4}, \emph{Am. J. Phys.} {\bf #1} (#3) #2}
\def\anp#1#2#3#4{\emph{#4}, \emph{ Ann. Phys.} {\bf #1} (#3) #2}
\def\sa#1#2#3#4{\emph{#4}, \emph{ Sov. Astro.} {\bf #1} (#3) #2}
\def\sia#1#2#3#4{\emph{#4}, \emph{ SIAM J. Sci. Statist. Comput.} {\bf #1} (#3) #2}
\def\aa#1#2#3#4{\emph{#4}, \emph{ Astron. Astrophys.} {\bf #1} (#3) #2}
\def\mnras#1#2#3#4{\emph{#4}, \emph{ Mon. Not. R. Astr. Soc.} {\bf #1} (#3) #2}
\def\npb#1#2#3#4{\emph{#4}, \emph{ Nucl. Phys. B} {\bf #1}, #2 (#3)}
\def\npa#1#2#3#4{\emph{#4}, \emph{ Nucl. Phys. A} {\bf #1} (#3) #2}

\def\prsla#1#2#3#4{\emph{#4}, \emph{ Proc. R. Soc. London, Ser. A} {\bf #1} (#3) #2}
\def\jhep#1#2#3#4{\emph{#4}, \emph{ JHEP} {\bf #1} (#2) #3}
\def\jcap#1#2#3#4{\emph{#4}, \emph{ JCAP} {\bf #1} (#2) #3}

\def\nuca#1#2#3#4{\emph{#4}, \emph{ Nuovo Cimento A } {\bf #1} (#3) #2}
\def\nucb#1#2#3#4{\emph{#4}, \emph{ Nuovo Cimento B } {\bf #1} (#3) #2}
\def\ijmp#1#2#3#4{\emph{#4}, \emph{ Int. J. Mod. Phys. D} {\bf #1} (#3) #2}
\def\atmp#1#2#3#4{\emph{#4}, \emph{ Adv. Theor. Math. Phys.} {\bf #1} (#3) #2}
\def\ptps#1#2#3#4{\emph{#4}, \emph{ Prog. Theor. Phys. Suppl.} {\bf #1} (#3) #2}
\def\ptp#1#2#3#4{\emph{#4}, \emph{ Prog. Theor. Phys.} {\bf #1} (#3) #2}
\def\lmp#1#2#3#4{\emph{#4}, \emph{ Lett. Math. Phys.} {\bf #1} (#3) #2}
\def\cpam#1#2#3#4{\emph{#4}, \emph{ Comm. Pure Appl. Math.}  {\bf #1} (#3) #2}
\def\adv#1#2#3#4{\emph{#4}, \emph{ Adv. Phys.}  {\bf #1} (#3) #2}
\def\zh#1#2#3#4{\emph{#4}, \emph{ Zh. Eksp. Teor. Fiz.}  {\bf #1} (#3) #2}
\def\mplb#1#2#3#4{\emph{#4}, \emph{ Mod. Phys. Lett. B} {\bf #1} (#3) #2}
\def\jams#1#2#3#4{\emph{#4}, \emph{ J. Austral. Math. Soc. B} {\bf #1} (#3) #2}
\def\appa#1#2#3#4{\emph{#4}, \emph{ Acta Phys. Polonica A} {\bf #1} (#3) #2}
\def\appb#1#2#3#4{\emph{#4}, \emph{ Acta Phys. Polonica B} {\bf #1} (#3) #2}

\def\nat#1#2#3#4{\emph{#4}, \emph{Nature} {\bf #1} #2 (#3)}
\def\natcom#1#2#3#4{\emph{#4}, \emph{Nature Commun.} {\bf #1} (#3) #2}
\def\natphys#1#2#3#4{\emph{#4}, \emph{Nature Physics} {\bf #1} (#3) #2}
\def\natmat#1#2#3#4{\emph{#4}, \emph{Nature Mat.} {\bf #1} (#3) #2}


\def\science#1#2#3#4{\emph{#4}, \emph{Science} {\bf #1} (#3) #2}
\def\sciadv#1#2#3#4{\emph{#4}, \emph{Sci. Adv.} {\bf #1} (#3) #2}

\def\arcmp#1#2#3#4{\emph{#4}, \emph{Annual Rev. of Cond. Matter Physics} {\bf #1} (#3) #2}
\def\zphys#1#2#3#4{\emph{#4}, \emph{Z. Phys.} {\bf #1}, (#3) #2}
\def\ncs#1#2#3#4{\emph{#4}, \emph{Nuovo Cimento Suppl.} {\bf #1} (#3) #2}
\def\physb#1#2#3#4{\emph{#4}, \emph{Physica B} {\bf #1}, (#3) #2}
\def\jpcm#1#2#3#4{\emph{#4}, \emph{J. Phys.: Condens. Matter } {\bf #1} (#3) #2}
\def\pnas#1#2#3#4{\emph{#4}, \emph{Proc. Nat. Academy Sciences} {\bf #1} (#3) #2}
\def\sssr#1#2#3#4{\emph{#4}, \emph{Izv. Akad Nauk SSSR, ser. fiz.} {\bf #1} (#3) #2}
\def\jpg#1#2#3#4{\emph{#4}, \emph{ J. Phys. G} {\bf #1} (#3) #2}
\def\chinpb#1#2#3#4{\emph{#4}, \emph{Chin. Phys. B} {\bf #1} (#3) #2}
\def\njp#1#2#3#4{\emph{#4}, \emph{ New J. Phys.} {\bf #1} (#3) #2}
\def\frontphys#1#2#3#4{\emph{#4}, \emph{ Front. Phys.} {\bf #1} (#3) #2}
\def\epl#1#2#3#4{\emph{#4}, \emph{ EPL} {\bf #1} (#3) #2}
\def\rmp#1#2#3#4{\emph{#4}, \emph{ Rev. Mod. Phys.} {\bf #1}, #2 (#3)}
\def\rpp#1#2#3#4{\emph{#4}, \emph{ Rep. Prog. Phys.} {\bf #1}, #2 (#3)}

\def\hepph#1#2{{ hep-ph }{#1} (#2)}
\def\arxiv#1#2#3{\emph{#3},{ arXiv }{#1} (#2)}
\def\hepth#1#2{{ hep-th }{#1} (#2)}
\def\grqc#1#2{{ gr-qc }{#1} (#2)}
\def\ibid#1#2#3#4{\emph{#4}, {\it ibid.} {\bf #1} (#3) #2}
\def\conphy#1#2#3#4{\emph{#4}, \emph{Contemporary Physics} {\bf #1}, (#3) #2}
\def\ppnp#1#2#3#4{\emph{#4}, \emph{ Prog. Part. Nucl. Phys} {\bf #1} (#3) #2}
\def\arnps#1#2#3#4{\emph{#4}, \emph{ Annu. Rev. Nucl. Part. Sci.} {\bf #1} (#3) #2}
\def\ijmpa#1#2#3#4{\emph{#4}, \emph{ Int. J. Mod. Phys. A} {\bf #1}, #2 (#3)}
\def\jams#1#2#3#4{\emph{#4}, \emph{ J. Austral. Math. Soc. B} {\bf #1} (#3) #2}
\def\appa#1#2#3#4{\emph{#4}, \emph{ Acta Phys. Polonica A} {\bf #1}, (#3) #2}
\def\nat#1#2#3#4{\emph{#4}, \emph{Nature} {\bf #1}, (#3) #2}
\def\science#1#2#3#4{\emph{#4}, \emph{Science} {\bf #1}, (#3) #2}
\def\arcmp#1#2#3#4{\emph{#4}, \emph{Annual Rev. of Cond. Matter Physics} {\bf #1}, (#3) #2}
\def\jcap#1#2#3#4{\emph{#4}, \emph{JCAP} {\bf #1}, (#3) #2}
\def\conphy#1#2#3#4{\emph{#4}, \emph{Contemporary Physics} {\bf #1}, (#3) #2}
\def\ptps#1#2#3#4{\emph{#4}, \emph{ Prog. Theor. Phys. Suppl.} {\bf #1} (#3) #2}
\def\ptp#1#2#3#4{\emph{#4}, \emph{ Prog. Theor. Phys.} {\bf #1} (#3) #2}
\def\apjsup#1#2#3#4{\emph{#4}, \emph{ Astrophys. J. Suppl. Ser.} {\bf #1} (#3) #2}
\def\eurphysjc#1#2#3#4{\emph{#4}, \emph{ Eur. Phys. J.  C} {\bf #1}, #2 (#3)}
\def\njp#1#2#3#4{\emph{#4}, \emph{ New J. Phys. } {\bf #1} (#3) #2}
\def\eurphysjplus#1#2#3#4{\emph{#4}, \emph{ Eur. Phys. J.  Plus} {\bf #1}, #2 (#3)}
%
\def\hepph#1#2{{ hep-ph }{#1} (#2)}
\def\hepth#1#2{{ hep-th }{#1} (#2)}
\def\astroph#1#2{{ astro-ph }{#1} (#2)}
\def\grqc#1#2{{ gr-qc }{#1} (#2)}
\def\ibid#1#2#3#4{\emph{#4}, {\it ibid.} {\bf #1} (#3) #2}

\def\contp#1#2#3#4{\emph{#4}, \emph{ Contemporary Physics} {\bf #1}, #2 (#3)}



%




\bibitem{bergstorm2000}
L. Bergstr\"om, \rpp{63}{793}{2000}{Non-baryonic dark matter: observational evidence and detection methods}.
\bibitem{bertone2005}
G. Bertone, D. Hooperb, and J. Silk, \prep{405}{279}{2005}{Particle dark matter: evidence, candidates and constraints}.
\bibitem{freese2013}
K. Freese, M. Lisanti, and C. Savage, \rmp{85}{1561}{2013}{Colloquium: Annual modulation of dark matter}.


\bibitem{ber98}
R. Bernabei {\it et al.}, \plb{424}{195}{1998}{Searching for WIMPs by the annual modulation signature}.
\bibitem{ber13}
R. Bernabei {\it et al.}, \eurphysjc{73}{2648}{2013}{Final model independent result of DAMA/LIBRA-phase1}.


\bibitem{cos18}
The COSINE-100 Collaboration, \nat{564}{83}{2018}{An experiment to search for dark-matter interactions using sodium iodide detectors}.




\bibitem{bartram2021}
C. Bartram et al. (ADMX Collaboration), \prl{127}{261803}{2021}{Search for Invisible Axion Dark Matter in the 3.3-4.2 $\mu$eV Mass Range}.

\bibitem{meng2021}
Yue Meng et al. (PandaX-4T Collaboration), \prl{127}{261802}{2021}{ Dark Matter Search Results from the PandaX-4T Commissioning Run}.

\bibitem{aprile2020}
E. Aprile et al. (XENON Collaboration), \prd{102}{072004}{2020}{Excess electronic recoil events in XENON1T}.

\bibitem{ber18}
G. Bertone and T.M.P. Tait, \nat{562}{51}{2018}{A new era in the search for dark matter}.








\bibitem{hochberg2016}
Y. Hochberg, T. Lin, and K.M. Zurek, \prd{94}{015019}{2016}{Detecting ultralight bosonic dark matter via absorption in superconductors}.

\bibitem{knapen2017}
S. Knapen, T. Lin, and K.M. Zurek, \prd{95}{056019}{2017}{Light dark matter in superfluid helium: Detection with multi-excitation production}.

\bibitem{knapen2018}
S. Knapen, T. Lin, M. Pyle, and K.M. Zurek, \plb{785}{386}{2018}{
Detection of light dark matter with optical phonons in polar materials}.

\bibitem{budnik2018}
R. Budnik, O. Cheshnovsky, O. Slone, and T. Volansky, \plb{782}{242}{2018}{Direct detection of light dark matter and solar neutrinos via color center production in crystals}.

\bibitem{hochberg2018}
Y. Hochberg, Y. Kahn, M. Lisanti, K.M. Zurek, A.G. Grushin, R. Ilan, S. M. Griffin, Z-F. Liu, S.F. Weber, and J.B. Neaton, \prd{97}{015004}{2018}{Detection of 
sub-MeV dark matter with three-dimensional Dirac materials}.

\bibitem{liang2018}
T. Liang, B. Zhu, R. Ding, and T. Li, \ijmpa{33}{1850135}{2018}{Direct detection of axion-like particles in Bismuth-based topological insulators}.

\bibitem{bar18}
M. Baryakhtar, J. Huang, and R. Lasenby, \prd{98}{035006}{2018}{Axion and hidden photon dark matter detection with multilayer optical haloscopes}.







\bibitem{mayet2016}
F. Mayet, A.M. Green, J.B.R. Battat, J. Billard, N. Bozorgnia,
G.B. Gelmini, P. Gondolo, B.J. Kavanagh, S.K. Lee, D. Loomba,
J. Monroe, B. Morgan, C.A.J. O’Hare, A.H.G. Peter, N.S. Phan, and
S.E. Vahsen, \prep{627}{1}{2016}{ A review of the discovery reach of directional Dark Matter detection}.

\bibitem{gaskins2016}
J. M. Gaskins, \contp{57}{496}{2016}{A review of indirect searches for particle dark matter}.

\bibitem{haldane2004}
F. D. M. Haldane, \prl{93}{206602}{2004}{Berry Curvature on the Fermi Surface: Anomalous Hall Effect as a Topological Fermi-Liquid Property}.

\bibitem{xiao2010}
Di Xiao, Ming-Che Chang, and Qian Niu, \rmp{82}{1959}{2010}{Berry phase effect on electronic properties}.

\bibitem{iwazaki2020}
A. Iwazaki, \plb{811}{135861}{2020}{A new method for detecting axion with cylindrical superconductor}.
\bibitem{iwazaki2021}
A Iwazaki, \npb{963}{115298}{2021} 
{\it Axion-radiation conversion by super and normal conductors}.
\bibitem{kishimoto2022}
Y. Kishimoto and K. Nakayama, \plb{827}{13}{2022}{Electric current on surface of a metal/superconductor in axion/hidden-photon background}.


\bibitem{kah22}
Y. Kahn and T. Lin, \rpp{65}{066901}{2022}{Searches for light dark matter using condensed matter systems}.



\bibitem{nak14}
L. Nakonieczny and M. Rogatko, \prd{90}{106004}{2014}{Analytic study on backreacting 
holographic superconductors with dark matter sector}.
\bibitem{nak15}
L. Nakonieczny, M. Rogatko, and K.I. Wysoki\'nski, \prd{91}{046007}{2015}{Magnetic field 
in holographic superconductors with dark matter sector}.
\bibitem{nak15a} L. Nakonieczny, M. Rogatko, and K.I. Wysoki\'nski, \prd{92}{066008}{2015}
{Analytic investigation of holographic phase transitions influenced by
dark matter sector}.
\bibitem{rog16} M. Rogatko and K.I. Wysoki\'nski, \jhep{03}{2016}{215}{P-wave holographic 
superconductor/insulator phase
transitions affected by dark matter sector}.
\bibitem{rog15a} 
M. Rogatko and K.I. Wysoki\'nski, \jhep{12}{2015}{041}{Holographic vortices in the 
presence of dark matter sector}.
\bibitem{rog16a}
M. Rogatko and K.I. Wysoki\'nski, \jhep{10}{2016}{152}{Condensate flow in holographic models in the presence of dark matter}.
\bibitem{rog17}
M. Rogatko and K.I. Wysoki\'nski, \prd{96}{026015}{2017}{Viscosity bound for anisotropic superfluids with dark matter sector}.
\bibitem{kic21}
B. Kiczek, M. Rogatko, and K. I. Wysoki\'nski, \jcap{01}{2021}{063}{Holographic DC SQUID in the presence of dark matter}.





\bibitem{pen15a}
Y. Peng, \plb{750}{420}{2015}{Holographic entanglement entropy in superconductor 
phase transition with dark matter sector}.
\bibitem{pen15b}
Y. Peng, Q. Pan, and Y. Liu, \npb{915}{69}{2017}{A general holographic insulator/superconductor model with dark matter sector away from the probe limit}.




\bibitem{ste12}
M. A. Stephanov and Y. Yin, \prl{109}{162001}{2012}{Chiral kinetic theory}.
\bibitem{bernevig2013} 
 B. Andrei Bernevig with Taylor L. Hughes, {\it Topological insulators and topological superconductors}, Princeton University Press (2013). 
\bibitem{fukushima2008}
K. Fukushima, D. E. Kharzeev, and H. J. Warringa, \prd{78}{074033}{2008}{Chiral magnetic effect}.

\bibitem{ste15}
M. Stepanov, H. -u. Yee, and Y. Yin, \prd{91}{125014}{2015}{Collective modes of chiral kinetic theory in a magnetic field}.
\bibitem{wan91}
X. -N. Wang, \prd{43}{104}{1991}{Role of multiple minijets in high-energy hadronic reactions}.



\bibitem{chen2014} 
J.-Y. Chen, D. T. Son, M. A. Stephanov, H.-U. Yee, and Y. Yin, \prl{113}{182302}{2014}{Lorentz Invariance in Chiral Kinetic Theory}.



\bibitem{dan18}
R. M. A. Dantas, F. Pena-Benitez, B. Roy, and P. Surowka, \jhep{12}{2018}{069}{Magnetotransport in multi-Weyl semimetals: a kinetic theory approach}.





\bibitem{guo20}
R. Guo, C. Liu, S. Mao, X. Xue, R. J. Long, and L. Zhang, \mnras{495}{4828}{2020}{Measuring the local dark matter density with LAMOST DR5 and Gaia DR2}.
\bibitem{loe22}
A. Loeb, \prd{105}{L0911903}{2022}{New way to limit the interaction of dark matter with baryons}.
\bibitem{beringer2012} 
J. Beringer, et al., \prd{86}{01001}{2012}{Review of particle physics}.
\bibitem{tho22}
A. W. Thomas, X. G. Wang, and A. G. Williams, \prl{129}{011807}{2022}{Sensitivity of parity-violating electron scattering to a dark photon}.




\end{thebibliography}
\end{document}